\documentclass[namedreferences]{SolarPhysics}
\usepackage[optionalrh]{spr-sola-addons} 
\usepackage{epsfig}          
\usepackage{graphicx}        
\usepackage{color}           
\usepackage{url}             
\newcommand{\citet}[1]{\inlinecite{#1}}  
\renewcommand{\cite}[1]{\opencite{#1}}   

\newcommand{\figspath}{.}

\newcommand{\Haouter}{\mbox{\Halpha\,-\,700\,m\AA}}
\newcommand{\HaDinner}{\mbox{\Halpha\,$\pm$\,350\,m\AA}}
\newcommand{\HaDouter}{\mbox{\Halpha\,$\pm$\,700\,m\AA}}
\newcommand{\is}{\!=\!}
\newcommand{\kms}{\hbox{km$\;$s$^{-1}$}}

\newcommand{\aspcs}{{\it Astron.\ Soc.\ Pacific Conf.\ Series}}

\def\ao{{\it Appl. Opt.}}

\def\CaII{\mbox{Ca\,\sc{ii}}} 
\def\CaIII{\mbox{Ca\,\sc{iii}}} 
 
\def\Halpha{\mbox{H\hspace{0.1ex}$\alpha$}}
\def\CaIIK{\mbox{Ca\,\sc{ii}\,\,K}}
\def\CaIIH{\mbox{Ca\,\sc{ii}\,\,H}}

\def\HK{\mbox{H~and~K}} 
\def\Kthree{\mbox{K$_{3}$}}
\def\Hthree{\mbox{H$_{3}$}}
\def\KtwoV{\mbox{K$_{2V}$}}
\def\HtwoV{\mbox{H$_{2V}$}}
\def\HtwoR{\mbox{H$_{2R}$}}

\newcommand{\etal}{{\it et al.}}
\newcommand{\eg}{{\it e.g.},}

\newcommand{\aap}{{\it Astron. Astrophys.}}

\newcommand{\apj}{{\it Astrophys. J.}}
\newcommand{\apjl}{   {\it Astrophys. J.}}

\newcommand{\solphys}{{\it Solar Phys.}}

\begin{document}
\begin{article}

\begin{opening}

\title{DOT Tomography of the Solar Atmosphere\\
       VII.  Chromospheric Response to Acoustic Events}
\author{R.J.~\surname{Rutten}$^{1,2}$\sep
        B.~\surname{van Veelen}$^1$\sep
        P.~\surname{S\"utterlin}$^{1}$
       }
\runningauthor{Rutten, van Veelen, S\"utterlin}
\runningtitle{Chromospheric Response to Acoustic Events}
\institute{$^1$Sterrekundig Instituut, Utrecht University, 
               The Netherlands\\
              email: \url{B.vanVeelen@astro.uu.nl}
              email: \url{R.J.Rutten@astro.uu.nl}
              email: \url{P.Suetterlin@astro.uu.nl}\\
           $^2$Institute of Theoretical Astrophysics, 
               University of Oslo, Norway}
              
\date{Received: 10 October 2007 / Accepted: 29 December 2008}

\begin{abstract}
  We use synchronous movies from the Dutch Open Telescope sampling the
  G band, Ca\,{\sc ii}\,H, and \Halpha{} with five-wavelength profile
  sampling to study the response of the chromosphere to acoustic
  events in the underlying photosphere.  We first compare the
  visibility of the chromosphere in Ca\,{\sc ii}\,H and \Halpha,
  demonstrate that studying the chromosphere requires \Halpha{} data,
  and summarize recent developments in understanding why this is so.
  We construct divergence and vorticity maps of the photospheric flow
  field from the G-band images and locate specific events through the
  appearance of bright Ca\,{\sc ii}\,H grains.  The reaction of the
  \Halpha{} chromosphere is diagnosed in terms of brightness and
  Doppler shift.  We show and discuss three particular cases in
  detail: a regular acoustic grain marking shock excitation by
  granular dynamics, a persistent flasher which probably marks
  magnetic-field concentration, and an exploding granule.  All three
  appear to buffet overlying fibrils, most clearly in
  Dopplergrams. Although our diagnostic displays to dissect these
  phenomena are unprecedentedly comprehensive, adding even more
  information (photospheric Doppler tomography and magnetograms,
  chromospheric imaging and Doppler mapping in the ultraviolet) is
  warranted.
\end{abstract}
\end{opening}

\section{Introduction}

``Acoustic events'' (or ``seismic events'') is the term used by
P.~Goode and coworkers 
 (\cite{1987SoPh..110..237S}; 
  \cite{1992ApJ...387..707G}; 
  \cite{1993ApJ...408L..57R}; 
  \cite{1995ApJ...444L.119R}; 
  \cite{1998ApJ...495L..27G}; 
  \cite{2000ApJ...535.1000S}) 
to describe localized small-scale happenings in the granulation that
produce excessive amounts of upward-propagating acoustic waves in the
upper photosphere, with the claim that these indicate the kinetic
sources of the global $p$-modes, ``the smoke from the fire exciting
the solar oscillations''
  (\cite{1995STIN...9610005G}). 
Their technique was to sample the Doppler modulation of a suitable
spectral line at different profile heights through Fabry-Perot imaging
with rapid wavelength shifting, and then isolate the surface locations
with the largest oscillatory amplitude at five-minute periodicity and
upward propagating phase.  Their result was that this measure
of ``acoustic flux''
tends to be maximum above intergranular lanes, in particular
those in which a small granule has vanished in so-called granular
collapse.
 
Acoustic excitation through granular dynamics was also addressed
theoretically and through numerical hydrodynamics simulations by
Rast
  (\citeyear{1995ApJ...443..863R}, 
   \citeyear{1999ApJ...524..462R}), 
  \citet{2000ApJ...535..464S} 
and
  \citet{2000ApJ...541..468S}, 
confirming the picture of small vanishing granules, especially at
mesogranular downdraft boundaries, acting as ``collapsars'' to excite
upward-propagating waves, in particular at three-minute periodicity
corresponding to the photospheric acoustic cut-off frequency.

In this paper we address the reaction of the overlying chromosphere to
such events.
  \citet{2000ApJ...541..468S} 
suggested that they may also cause \CaII\
\HtwoV\ and \KtwoV\ cell grains,
an enigma during many decades (see review by
  \cite{1991SoPh..134...15R}) 
that was eventually solved by 
  Carlsson and Stein 
  (\citeyear{Carlsson+Stein1994}, 
   \citeyear{1997ApJ...481..500C}) 
who identified them as marking upward-propagating acoustic shocks.  The
announcement by
  \citet{2002AAS...200.5305G} 
that indeed acoustic events cause such acoustic grains in the
chromosphere was followed up by
  \citet{2002A&A...390..681H} 
who found that, while extreme acoustic events do tend to correlate with
the subsequent appearance of exceptionally bright \CaII\ \KtwoV\
internetwork grains, this correspondence is far from a one-to-one
correlation.

\begin{figure}
  \centering
  \includegraphics[width=\textwidth]{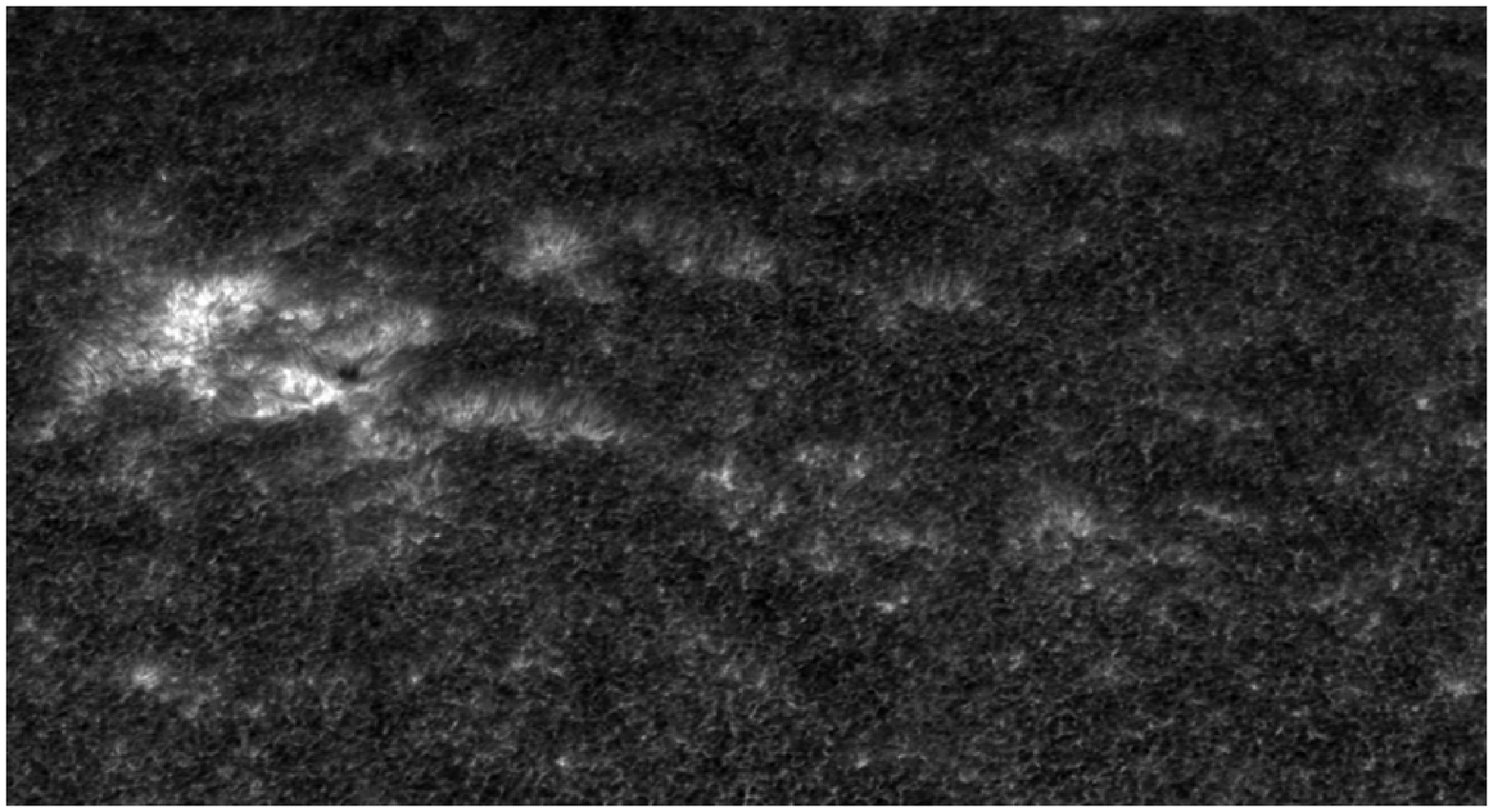}\\[1mm]
  \includegraphics[width=\textwidth]{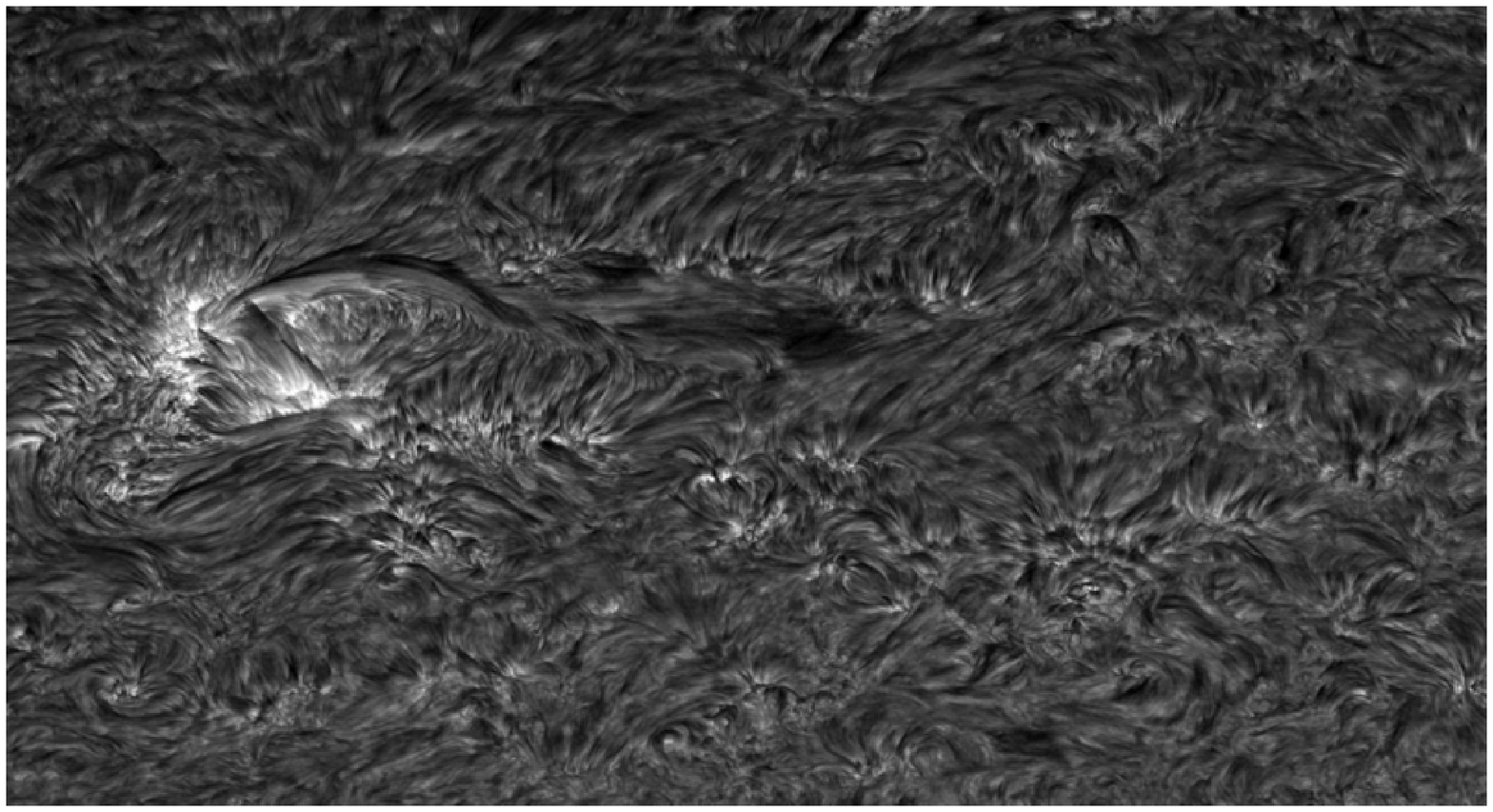}
  \caption[]{
  Simultaneous \CaIIH\ and \Halpha\ image mosaics taken with the Dutch
  Open Telescope on 4 October 2005.  The field of view is close to
  the limb (located off the top) and measures about $265\times
  143$~arcsec$^2$.  Spectral passbands: FWHM 1.4\,\AA\ for \CaIIH,
  0.25\,\AA\ for \Halpha.  The same area was observed with the SST and
  described by
   \citet{2006ApJ...648L..67V}. 
  We invite the reader to enlarge this and the following figures
  on-screen.
%
  }\label{fig:mosaics}
\end{figure}

\begin{figure}
  \centering
  \includegraphics[width=\textwidth]{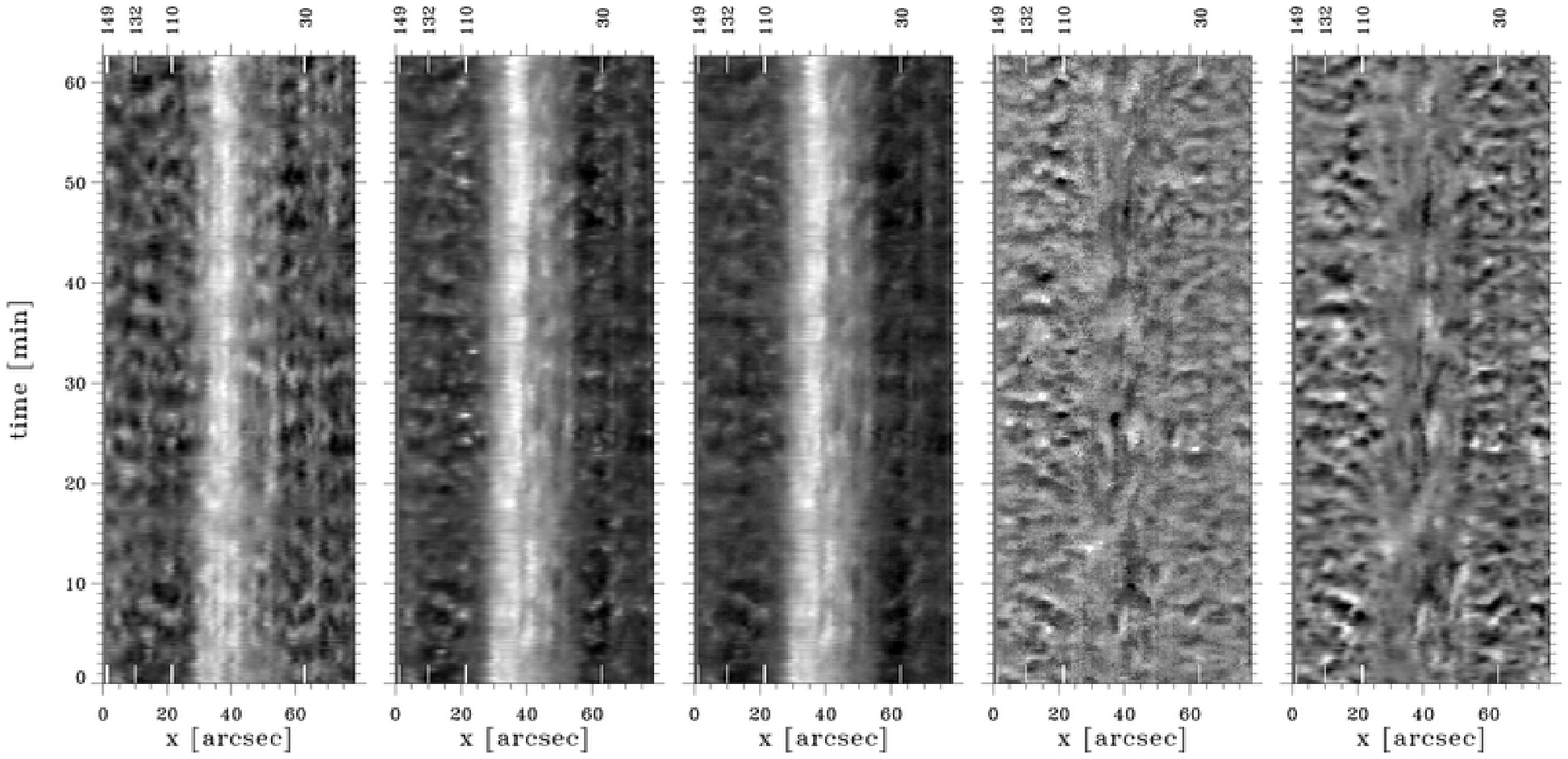}
  \caption[]{
  Various diagnostic measures of \CaIIH\ plotted as space\,-\,ime
  ``timeslice'' evolution plots, assembled by B.W.~Lites from a
  spectral sequence taken by himself and W.~Kalkofen in 1984 with the Dunn
  Solar Telescope at the National Solar Observatory/Sacramento Peak.
  The waves in the network part (bright strip at the center) were 
  analyzed by
  \citet{1993ApJ...414..345L}. 
  The waves in the internetwork columns identified by numbers along the top
  were numerically simulated by
  \citet{1997ApJ...481..500C}. 
  The spatial coordinate is measured along the slit of the
  spectrograph.  The first three panels have logarithmic greyscales to
  reduce the large contrast between network and internetwork.
  Horizontal greyish ``erasures'' are due to lesser seeing.  First
  panel: intensity integrated over a spectral band of 0.9\,\AA\ width
  centered on line center.  Second panel: the same for 0.16\,\AA\
  bandpass.  Third panel: intensity of the core-profile minimum.
  Fourth panel: wavelength variation of the core-profile minimum.
  Fifth panel: Dopplergram ratio $(R-V)/(R+V)$ with $R$ and $V$ the
  intensities in 0.16\,\AA\ passbands at \HtwoR\ and \HtwoV.
  }\label{fig:lites}
\end{figure}

We now turn to \Halpha\ rather than \HK\ as diagnostic of the
chromospheric response to acoustic events.  It is important to note
and to explain that \Halpha\ is a much better chromospheric diagnostic
than the \CaII\ \HK\ lines are.  This is not the case for simple
Saha-Boltzmann LTE partitioning in which \HK\ have larger opacity
than
\Halpha\ throughout the atmosphere
  (see Figure~6 of \cite{2006A&A...449.1209L} 
and the second student exercise at 
\url{http://www.astro.uu.nl/~rutten/education/rjr-material/ssa}),
and it is also not the case in the standard NLTE statistical and
hydrostatic equilibrium VAL modeling of
  Vernazza, Avrett, and Loeser (1973, 1976, 1981)
      \nocite{VALI} \nocite{VALII} \nocite{VALIII}
%
in which the line center of \CaIIK\ is formed higher than the line
center of \Halpha\ (see the celebrated Figure~1 of \cite{VALIII}).  We
use displays from older data here to argue that the real Sun does not
conform.

Figure~\ref{fig:mosaics} compares Dutch Open Telescope (DOT) images
using the two diagnostics.  The upper image is taken in Ca\,II\,H and
shows a mixture of mid- and upper-photospheric contributions (reversed
granulation and acoustic grains) and chromospheric contributions
(bright plage, network, and straws in hedge rows as	 described by
 \cite{2007ASPC..368...27R}). 
The lower panel shows the same area observed in H$\alpha$ line center.
Comparison of the two demonstrates unequivocally that in such filter
imaging H$\alpha$ presents a much more complete picture of the
chromosphere.  In our opinion, the chromosphere is best defined as the
collective of ubiquitous fibrils seen in this line.  Most of these are
internetwork-spanning structures, which probably outline magnetic
canopies.  Only truly quiet internetwork areas (there are some in the
lower part of the image) are not masked off by such canopy fibrils but
show much shorter, highly dynamic loops and grainy brightness patterns
that are not connected to network
  (see also \cite{2007ApJ...660L.169R}). 
None of these structures is observed in the upper panel,
except for the straws which correspond to bright \Halpha\ fibril
endings.

What about the spectral passbands in this comparison?  The Doppler cores of
\HK\ are much narrower than for \Halpha, but \HK\ filter
bandpasses are usually wider: FWHM 3\,\AA\ for {\em Hinode\/}, 1.4\,\AA\ for
the DOT, 0.6\,\AA\ for the Lyot filter at the German VTT 
  (\eg\ \cite{2007A&A...462..303T}), 
0.3\,\AA\ for the Lockheed-Martin filter at the former SVST
  (\cite{Brandt+Rutten+Shine+Trujillo1992};
   \cite{1999ApJ...517.1013L}), 
and also 0.3\,\AA\ for the Halle filter at the NSO/SP/DST 
 (\eg\ \cite{2004ApJ...604..936B}, 
who assigned propagation speeds and mode conversions to chromospheric
oscillations naively adopting the VAL3C line-center height
difference).  Such \HK\ filter imaging adds considerable inner-wing
contributions from the upper photosphere, including bright reversed
granulation and yet brighter acoustic grains which may wash out the
line-center-only contribution from chromospheric fibrils in the
internetwork except where these are very bright, as in straws.  The 
narrowest-band images with high spatial resolution published so far are
R.A.~Shine's 0.3\,\AA\ ones in
  \citet{1999ApJ...517.1013L}, 
%
which only show reversed granulation and acoustic grains in the
internetwork.  Narrow-band \Kthree\ spectroheliograms as the 0.1\,\AA\
ones in the collection of
  \citet{Title1966a} 
%
have lower angular resolution but do not suggest the presence of
similar masses of internetwork \CaIIK\ fibrils as seen in \Halpha\ at
similar resolution, except near active network and plage.

Figure~\ref{fig:lites} extends the \CaIIH\ passband comparisons 
in Figure~1 of
  \citet{1999ASPC..183..383R}
 and Figure~9 of
  \citet{2001A&A...379.1052K} 
with additional spectral measures from the ancient but high-quality
\CaIIH\ spectrogram sequence of
  \citet{1993ApJ...414..345L} 
also used by
   \citet{1997ApJ...481..500C}. 
The first three panels show that narrower \CaIIH\ passband produces
larger network-to-internetwork brightness contrast, but does not
significantly change the morphology of the observed scene.  The
oscillation patterns in the internetwork, brightest at \HtwoV, remain
visible even in the line-core intensity in the third panel. The
line-core shift in the fourth panel confirms the acoustic nature of
these patterns.  Thus, even at full spectral resolution the \CaIIH\
core in this internetwork sample is dominated by radial three-minute waves,
without obliteration by overlying fibrils or fibril-aligned
motions. The final panel showing the
\HtwoR$-$\HtwoV\ Dopplergram ratio strengthens this conclusion.  This
measure comes closest to exhibiting chromospheric fibrilar structure
in the form of oscillatory branches jutting out from the network with
time, but regular three-minute oscillation patterns dominate further
away in the internetwork.
 

The lower-left panel of Figure~21 of
  \citet{2001A&A...379.1052K} 
demonstrated the close correspondence between \CaIIH\ line-center
dynamics and underlying photospheric dynamics in the internetwork
parts of these spectrograms in Fourier terms.  In fact, it should be
strange if the line-center formation layer would not fully partake in
the shock dynamics in these data because substantial in-phase
upper-atmosphere downdraft, above the next upcoming grain-causing
acoustic shock, is required to give the
\HtwoV\ and \KtwoV\ grains their characteristic spectral asymmetry
as shown in the formation breakdown diagrams of
   \citet{1997ApJ...481..500C}. 
They actually obtained their best sequence reproduction for
near-network column 110 in Figure~\ref{fig:lites}, suggesting that
even there the field was too weak to upset \Hthree\ acoustic shock signatures. 
 
Thus, these well-studied \CaIIH\ spectrograms contradict the
presence of opaque chromospheric fibrils masking
upper-photosphere (or ``clapotispheric'') internetwork dynamics in
\HK, whereas \Halpha\ canopy fibrils ubiquitously do so.
The notion that the
\CaII\ core generally forms higher than the \Halpha\ core, or that 
chromospheric structures should generally be more opaque in \HK\ than
in \Halpha, seems questionable.  The observations suggest instead that
the fibrils that constitute the \Halpha\ chromosphere are either
transparent in \HK\ or, when opaque, are either pummeled by shocks
from below without hindrance (such as magnetic tension) or
represent a featureless, very dark blanket located well above the
clapotisphere.  

In summary, we suspect that much of the \Kthree\ internetwork dynamics is
dominated by shock modulation at heights around 1000~km, well below
the VAL3C \Kthree\ formation height of 1800\,--\,2000\,km and far
below most \Halpha\ canopy fibrils.

This large discrepancy with equilibrium modeling may become
understandable with the recent simulation of
 \citet{2007A&A...473..625L} 
implementing non-instantaneous hydrogen ionization and recombination.
Three aspects combine in such an explanation.  Firstly, the simulated
chromosphere is pervaded by shocks.  Secondly, the large difference
between hydrogen ionization/recombination balancing speed in hot
shocks and in cool post-shock gas implies that instantaneous
statistical equilibrium does not apply to the latter.
  \citet{2002ApJ...572..626C} 
already explained why this is so.  The large 10~eV excitation energy
of its $n\is2$ level causes hydrogen to be ionized fast in shocks but
to recombine slowly in the cool inter-shock phases so that the
ionization degree remains high also in the latter.  Thirdly, the
population of \Halpha's lower $n\is2$ level is closely coupled to the
ion population
 (through the Balmer continuum; see also 
  \cite{1994IAUS..154..309R}) 
and therefore remains similarly high at low post-shock temperature.
Thus, the large $n\is2$ excitation energy, being the very reason why
\Halpha\ has very small cool-gas opacity in Boltzmann balancing, slows
down the hydrogen recombination in post-shock cool gas so much that
its \Halpha\ opacity exceeds the LTE value by many orders of
magnitude.  Therefore, fibrils can be opaque in \Halpha\ whether they
are shock-hot or post-shock cool, and they even may be more opaque in
\Halpha\ than in \HK\ at any temperature -- in utter conflict
with equilibrium modeling.

The chromospheric non-equilibrium balancing between \CaII\ (12~eV
ionization energy) and \CaIII\ is probably faster (no top-heavy term
diagram) but has not yet been analyzed in similar detail; it is of
interest to evaluate the opacity ratio between \HK\ and \Halpha\ in
dynamic conditions.  Another point of interest is that chromospheric
fibrils seem to be more obvious in
\CaII~8542\,\AA\ than in \Hthree\ and \Kthree,
possibly due to larger Doppler sensitivity
 (\cite{2007arXiv0709.2417C}). 

Thus, \Halpha\ is the key diagnostic of the internetwork chromosphere.
Recent analyses of the chromosphere observed at high cadence in
\Halpha\ line-center with the Swedish 1-m Solar Telescope (SST) have 
shed new light on chromospheric fibrils,
showing that near network and plage ``dynamic fibrils'' tend to have
repetitive mass loading at three\,--\,five-minute shock periodicity 
  (\cite{2007ApJ...655..624D}) 
%
whereas the little loops in quiet-sun internetwork are even more
dynamic, presumably due to stronger buffeting by shocks where the
magnetic field is weak
  (\cite{2007ApJ...660L.169R}). 

In this paper we use image sequences from the Dutch Open Telescope
(DOT) with lower cadence than \Halpha\ really requires
   (\cite{2006ApJ...648L..67V}), 
but with \Halpha\ profile sampling and
synchronous co-spatial imaging in other wavelengths including
\CaIIH.  They permit us to isolate specific events in the photosphere
and inspect the chromospheric response in the apparent brightness of
\CaIIH\ and \Halpha, and particularly in \Halpha\ Doppler modulation.  We do
this here by presenting three exemplary cases: an acoustic grain, a
persistent flasher, and an exploding granule.

\begin{figure}
  \centering \includegraphics[width=100mm]{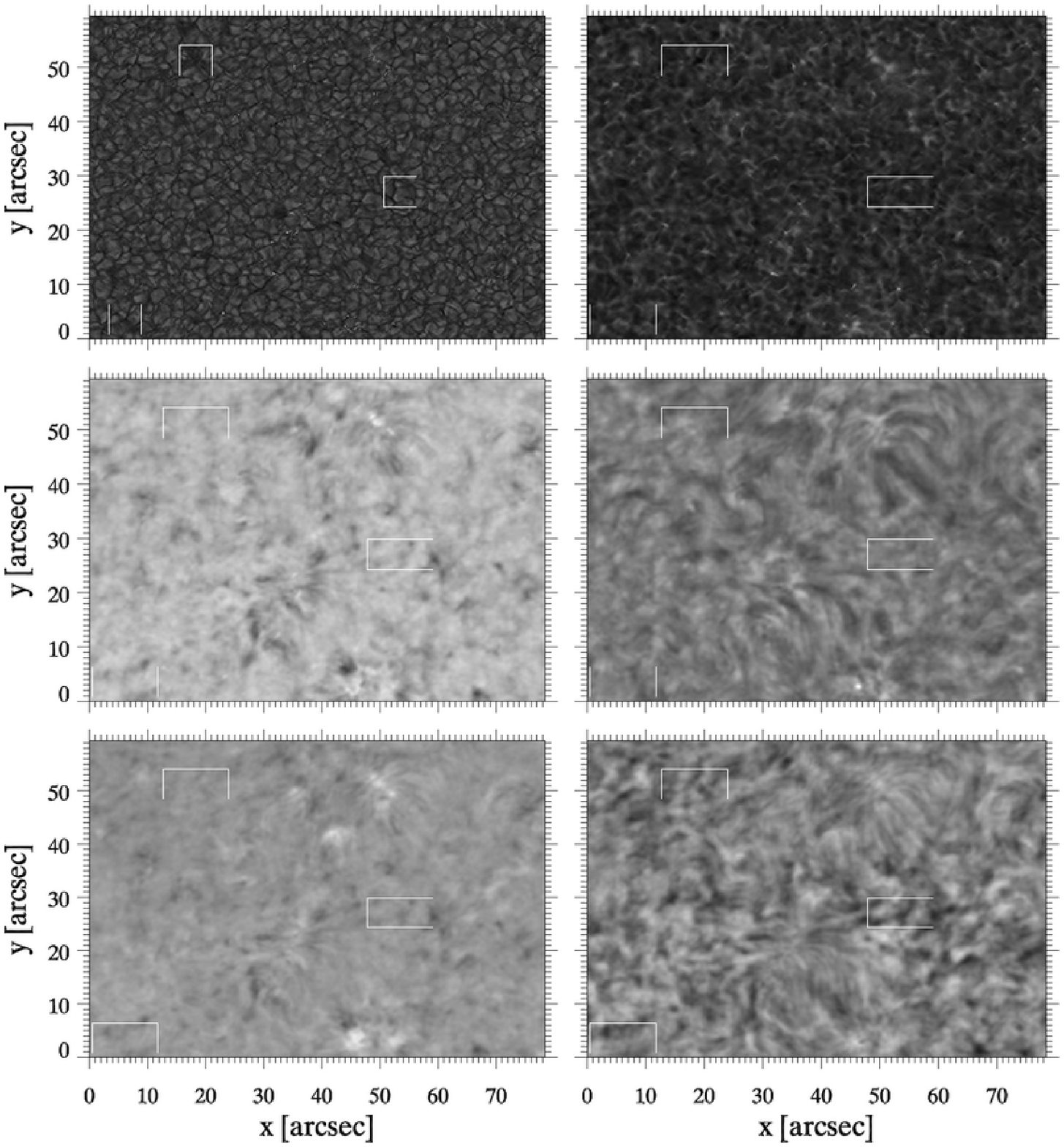} 
  \caption[]{
  Simultaneous sample images.  In row order: G band, \CaIIH\ line
  center, \Haouter, \Halpha\ line center, and HaDouter\ and
  \HaDinner\ Dopplergrams with bright implying downdraft.  The G-band
  image is Fourier filtered to pass only subsonic components.  The
  \Halpha\ images and Dopplergrams are filtered to pass only
  supersonic components.  The white boxes specify the three subfields
  analyzed in Figures \ref{fig:grain}\,--\,\ref{fig:exploder}, from left to
  right respectively.  The cutout boxes are larger for \CaIIH\ and
  \Halpha\ to show more spatial context in the timeslices of
  Figures \ref{fig:grain}\,--\,\ref{fig:exploder}.
  }\label{fig:samples}
\end{figure}

\section{Observations and Reduction}

The image sequences used here were obtained with the Dutch Open
Telescope (DOT) on La Palma on 14 October 2005 during
10:15:43\,--\,10:30:42~UT.  The telescope, instrumentation, and
data processing are described by
  \citet{2004A&A...413.1183R}. 

Synchronous and co-spatial images were taken in blue and red continuum
passbands, the G band around 4305\,\AA, \CaIIH\ alternatingly at line
center and at $\Delta \lambda \is -1$\,\AA, and at five \Halpha\ 
wavelengths: line center,
the \HaDinner\ inner-wing pair, and the \HaDouter\ outer-wing pair.
The seeing had mean Fried parameter $r_0 = 7$~cm for the G band, just
about good enough that the standard DOT speckle reconstruction
produced angular resolution close to the diffraction limit of the
45-cm aperture (0.2~arcsec in the blue).  The field of view measured
$79\times60$~arcsec$^2$ ($56\times43$~Mm$^2$), covering a very quiet
area at disk center containing only few and sparse weak-network clusters of
magnetic elements (bright points in the G band, bright grain clusters
in \CaIIH) as evidenced by the first two sample images in
Figure~\ref{fig:samples}.  This area was sufficiently quiet that
\Halpha\ shows only short, highly dynamic fibrils 
concentrated around the magnetic clusters.

The image sequences had slightly irregular cadence and were
interpolated to 30-image sequences at strict 30-second cadence.
The G-band images were subsonically Fourier-filtered in ($k, \omega$)
space, passing only components with apparent motion below the sound
speed of 7~\kms\ to remove the brightness modulation due to the
$p$-mode oscillations since in this study granular dynamics is
emphasized as pistoning agent.  Horizontal flows were measured using
the cross-correlation tracking algorithm of
  \citet{1986ApOpt..25..392N} 
and
  \citet{1988ApJ...333..427N}, 
using Gaussian FWHM widths of 1.5~arcsec and 330~seconds as boxcar averaging
parameters.  Differences between neighboring pixels were used to
evaluate derivatives to determine the local horizontal divergence and
vorticity in the flows.

The \Halpha\ sequences were also cone-filtered in ($k, \omega$) space
but in this case reversely, only passing the supersonic components
because here the oscillatory response of the chromosphere is of
interest.  Our intention was to remove stable fibrils in order to
enhance the visibility of oscillatory modulations such as spreading
rings, but in this very quiet area the fibrils are so dynamic that the
differences between the filtered and unfiltered sequences are small.
Finally, \Halpha\ Dopplergrams were constructed from the filtered
\HaDouter\ and \HaDinner\ image pairs.


We used a live multi-panel on-screen display similar to
Figures \ref{fig:grain}\,--\,\ref{fig:exploder} to inspect the various
diagnostics simultaneously and in mutual correspondence at large local
magnification, varying both the location and time coordinates of the
point of scrutiny within these image sequences.  The three cases
discussed below have been selected in this subjective manner as being
the most informative -- they are not ``typical'' but ``best''
examples.  However, since the field of view and the sequence duration
were small, they are nevertheless indicative of common occurrences on
the solar surface.  Automation of such a vision-based selection
process for larger data sets is nontrivial.

Figure~\ref{fig:samples} shows that all three examples lie in pure
internetwork without magnetic elements and with grainy rather than
fibrilar \Halpha\ fine structure.  They sample the thinnest and 
lowest sort of \Halpha\ chromosphere.

\section{Acoustic Grains}

\begin{figure}
  \centering
  \includegraphics[width=\textwidth]{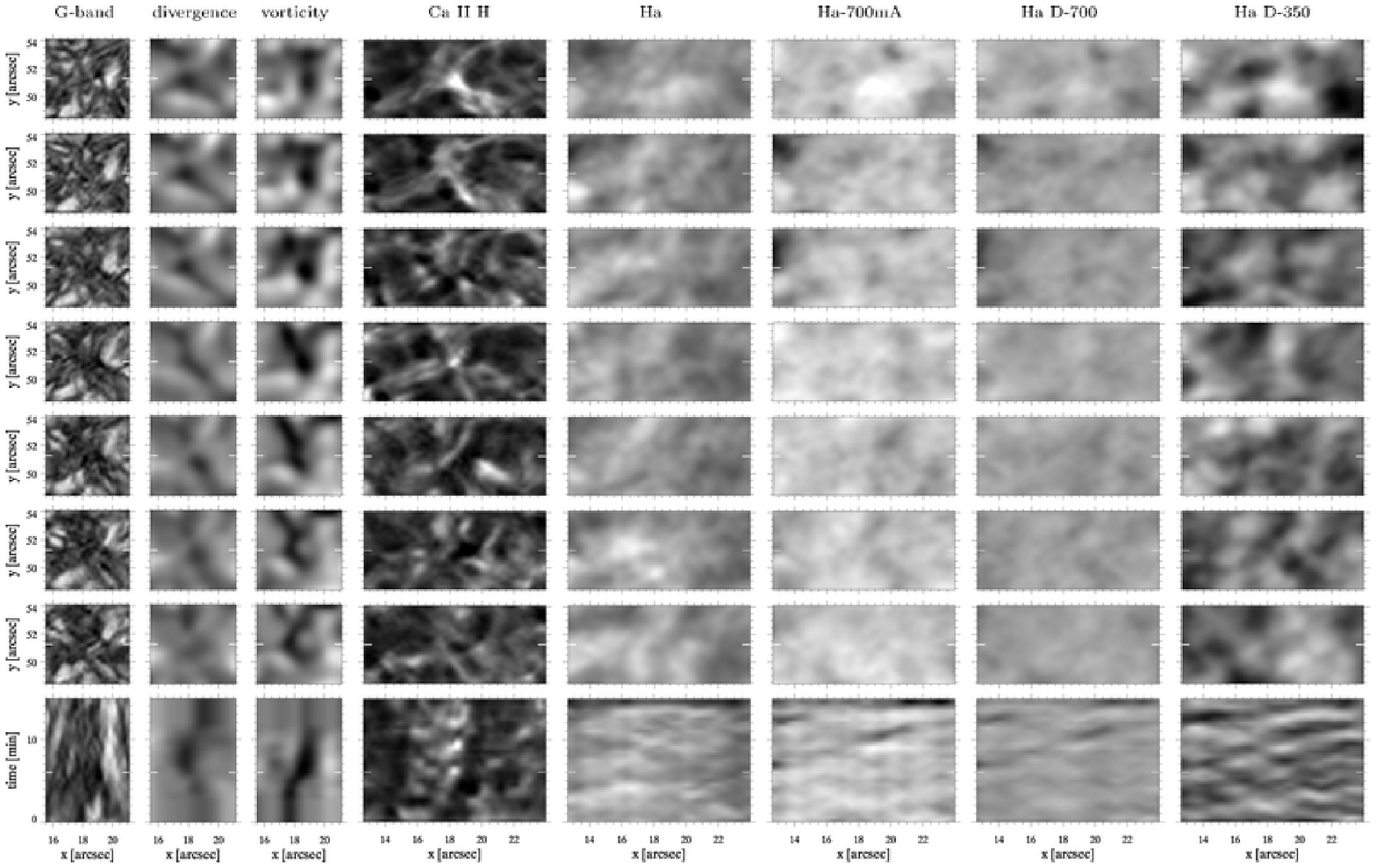}
  \caption[]{
  Acoustic \HtwoV\ grain.  The first seven rows are small image
  cutouts, in this case corresponding to the leftmost box in
  Figure~\ref{fig:samples}.  Each panel is spatially centered on the
  acoustic grain.  The diagnostics are specified at the top; the last
  two columns are the \HaDouter\ and \HaDinner\ Dopplergrams.  The
  cutouts are wider for the upper-atmosphere diagnostics (\CaIIH\ and
  \Halpha) in order to show more context in their $x-t$ timeslices at
  the bottom.  The latter show the brightness evolution along the
  horizontal cut through the center of the subfield defined by the
  white markers in each image cutout.  The time step between the
  consecutive image rows is one~minute, with time increasing upward to
  correspond with the time direction of the slices.  The image cutout
  sequences span seven minutes centered on the first appearance of the grain
  (bright small \CaIIH\ blob in the fourth column and fourth row).  In
  the second column bright and dark measure divergence and
  convergence, respectively.  In the third column bright implies
  counter-clockwise rotation.  In the Dopplergrams (last two columns)
  bright implies downdraft.  The grey scaling is the same per column
  throughout Figures~\ref{fig:grain}\,--\,\ref{fig:exploder} to enable
  intercomparison.  The timeslices have been temporally interpolated
  to 15-second cadence to gain display height.
  }\label{fig:grain} 
\end{figure}

Figure~\ref{fig:grain} shows the appearance of a regular \CaII\
\HtwoV\ ``grain train.''  The caption specifies the complex figure
layout which also holds for Figures~\ref{fig:patch} and
\ref{fig:exploder}.  The cutout image sequences (time increasing
upward in compliance with the $x\!-\!t$ timeslices at the bottom) are
centered on the sudden appearance of a \CaIIH\ grain (fourth row,
fourth column).  It re-appeared three and six~minutes later, marking a regular
\HtwoV\ shock sequence much like the triple grain occurrence at the
top of column 30 in the first panel of  Figure~\ref{fig:lites},
shown also in Figures~9 and 10 of
  \citet{2001A&A...379.1052K} 
and Fourier-analyzed there in detail, noting that such clear three-minute
periodicity requires absence of large five-minutes modulation but spatial
coincidence with the slower-evolving reversed-granulation background
pattern.  The latter was subsequently analyzed by
  \citet{2004A&A...416..333R}, 
  \citet{2005A&A...431..687L}, 
and
  \citet{2007A&A...461.1163C}. 
Such spatial coincidence is indeed seen to be the case in the
\CaIIH\ cutouts and timeslice in Figure~\ref{fig:grain}.  (Note that 
reversed granulation is not seen in \Halpha\ which has
insufficient opacity in the middle photosphere.)  These
three grains undoubtedly mark acoustic shocks as the ones simulated by
  \citet{1997ApJ...481..500C} 
including the column-30 ones of Figure~\ref{fig:lites} (their Figure~17,
Plate 21).

Do these \CaIIH\ grains mark an acoustic event?  If so, one expects
that a small granular feature disappeared about two minutes before the
grain train onset.  The corresponding area does show an intergranular
space in the wake of a bright granule that split apart and disappeared
five\,--\,two minutes earlier (G-band timeslice, not fully covered by the cutout
sequence).  Before the grain appeared there was weak divergence meeting
weak convergence at the splitting-granule site, eventually followed by
stronger convergence which bordered marked vorticity (dark blobs in the
upper panels of the second and third columns) which earlier was located
in the lane besides the splitting granule (third timeslice).
Together, the photospheric diagnostics suggest that these grains mark
acoustic excitation by a small whirlpool downdraft.

What happened in the overlying chromosphere?  This subfield is located,
as are the other two, in quiet internetwork without clear \Halpha\
fibrils.  The \Halpha\ line-center column shows nothing special at the
grain location.  The top \Haouter\ panel shows marked brightening
there, followed within two minutes by yet stronger darkening (seen the
timeslice, which extends longer).  The same pattern appears in the
\HaDouter\ Doppler timeslice so this dark-after-bright pattern is
mostly Doppler modulation, shifting the line out and into the
passband.  Due to the curvature of the line profile such Doppler
darkening tends to be stronger than counter-phase brightening, as is
the case here.  The same bright-dark pair is also recognizable in the
\HaDinner\ timeslice, more clearly preceded by another, smaller
bright-dark pair at the grain location, and with apparent connectivity
to longer features towards the top of the timeslice (beyond the top
cutout images), extending over four to five arcseconds.  Comparison with the
\HaDouter\ timeslice shows that these are weakly present there as
well.  Their extent might suggest wide wave spreading with height, but
the Dopplergram cutouts do not show corresponding rings around the
grain location.  Either the spreading wave ran into a finely
structured atmosphere, or made already-existing \Halpha\ fibrils shake
over extended lengths through shock buffeting.  If so, these fibrils
are not very distinct in \Halpha\ line center (also not in the
unfiltered images) and obtain their clearest mark of presence through
Doppler modulation.

Note that the simultaneous appearance of the second \CaIIH\ grain and
the marked \Haouter\ downdraft in the top row fits the scenario for
spectrally asymmetric grain formation of
  \citet{1997ApJ...481..500C}. 
Comparison of the \CaIIH\ and \HaDinner\ timeslices confirms that all
three \CaIIH\ grains indeed had concurrent chromospheric downdrafts.

\section{Persistent Flasher}

\begin{figure}
  \centering
  \includegraphics[width=\textwidth]{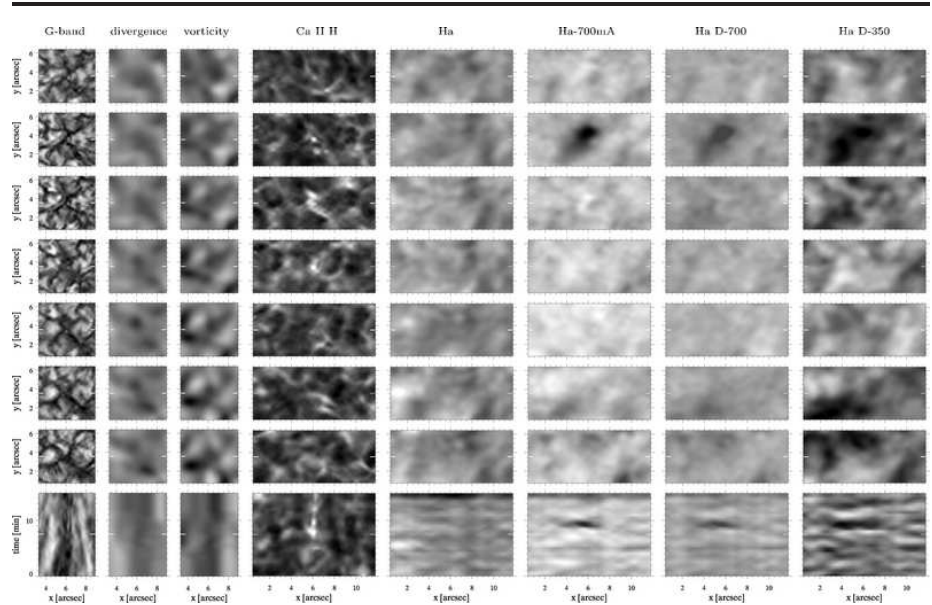}
  \caption[]{
  Persistent \HtwoV\ flasher which likely marks a magnetic patch, in
  the format of Figure~\ref{fig:grain}.  The cutouts correspond to the
  uppermost box in Figure~\ref{fig:samples}.  The mid-moment sampled by
  the fourth-row panels corresponds to the sudden appearance of a
  persistent flasher in \CaIIH\ (fourth column).
  }\label{fig:patch}
\end{figure}

Figure~\ref{fig:patch} is again centered on the first appearance of a
bright repetitive \CaIIH\ grain, but this one had continuous
brightness underlying its repeated brightening and was accompanied by a
small bright point in the G-band images and timeslice.  This behavior
suggests that it was a ``persistent flasher'' marking the presence of a
small internetwork magnetic flux concentration
  (Brandt \etal\ 1992, 1994).
     \nocite{Brandt+Rutten+Shine+Trujillo1992} 
     \nocite{Brandt+Rutten+Shine+Trujillo-Bueno1994} 
These are more easily located in \CaIIH\ image sequences
than as G-band bright
points and tend to come and go in the form of intermittent ``magnetic
patches'', showing up as bright grains when squeezed together and
vanishing when their concentration becomes less dense as imposed by
the granular flows advecting their field
  (\cite{2005A&A...441.1183D}). 
Neither the \CaIIH\ flasher nor the G-band bright point is seen in the
bottom half of the timeslices, indicating that the concentration was
initially more spread-out.  The second timeslice indeed shows persistent
convergence which is already evident from the squeezing together of
the large granules in the first timeslice.  The largest convergence is
in the neighboring lanes, up to the bright-point appearance (lower
cutouts in the second column).  Another speculation might be the
incidence of fluxtube collapse, but the convergence should then
maximize at the bright-point site.

The chromosphere shows marked response in the \Haouter\ and Doppler
columns: a dark blob in the second row which again is seen to result
from an up-down dark-bright Doppler stroke in the timeslices, most
clearly in the \HaDinner\ Doppler slice in the last column (partially
mimicked at lower contrast by the \Halpha\ line-center intensity in
the fifth slice), but in this case preceded by a brief bright
downstroke.  The size of the oscillatory blob is much larger than the
\CaIIH\ grains, especially in the second row, again suggesting
response by extended fibrilar structures.  The second row also
indicates strong updraft above the (very sharp) \CaIIH\ grain,
suggesting another mechanism than acoustic \HtwoV\ grain formation.
Fibril loading from the granular squeezing seen in the first
timeslice, sending gas upward along the field concentration indicated
by the \CaIIH\ flasher, may occur here following the scenario
suggested by
  \citet{1955ApJ...121..349B}. 
%

\section{Exploding Granule}

\begin{figure}
  \centering
  \includegraphics[width=\textwidth]{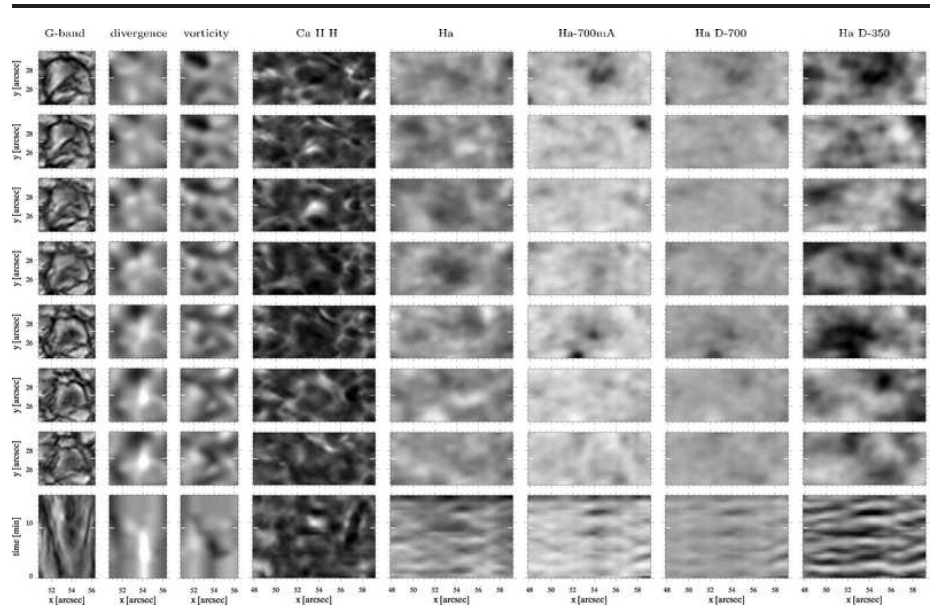}
  \caption[]{
  Exploding granule in the format of Figure~\ref{fig:grain},
  corresponding to the rightmost box in Figure~\ref{fig:samples}.  The
  mid-moment sampled by the fourth-row panels corresponds to the end
  of marked divergence (second column).
  }\label{fig:exploder} 
\end{figure}

Our third example in Figure~\ref{fig:exploder} concerns an exploding
granule to assess the suggestion of
  \citet{1995ApJ...443..863R} 
that such features may act as acoustic sources.  The cutout sequences
are centered on the moment at which the marked divergence in the
second timeslice ends.  The granule had already a large dark center
and was about to become three small granules with some shards in
between.  A minute later, the \CaIIH\ line became very bright at that
location, and after two more minutes very dark (top row).  The same
dark blob is seen along the top row in the \Halpha\ wing columns,
implying large updraft.  The \HaDinner\ Doppler timeslice in the last
panel again shows much more spatially extended oscillatory behavior
than the other diagnostics, suggesting fibril buffeting.  The \Halpha\
line-center intensity timeslice again mimics the Doppler behavior
partially.  The exploding granule seems to cause oscillation amplitude
increase above it which modulates the \Halpha\ core intensity.

\section{Discussion}

Each example shows large morphological difference between the
low-photosphere scene in the G-band, the high-photosphere scene in
\CaIIH, and the chromospheric scene in \Halpha.  The \Halpha\ cutouts do not
contain internetwork-spanning fibrils in this quiet region; each of
the three events therefore shows chromospheric response that would
otherwise have been blocked by overlying canopy fibrils.  The
responses are not very striking; in no case may one locate the
photospheric happening uniquely from its subsequent \Halpha\
signature.

The \Halpha\ chromosphere sampled by the image cutouts comes close to
our tentative description above of the \CaII\ \Hthree\ and \Kthree\
internetwork chromosphere as a featureless opaque blanket pummeled
from below.  They may indeed be about the same, as suggested by the
\Halpha\ downdrafts in Figure~\ref{fig:grain} above grains requiring
\Hthree\ downdraft for
\HtwoV\ profile asymmetry.  The much smaller temperature sensitivity of
the \HK\ opacity then indeed diminishes the \HK\ feature contrast.
Thanks to being an excited and narrower line, Ca\,II\,8542 combines
larger temperature and Doppler sensitivity to such pummeling but at
smaller blanket opacity.

Fibrilar structuring appears most clearly in the \Halpha\ Dopplergram
timeslices.  Our impression is that indeed this quiet \Halpha\
chromosphere is continuously pummeled by internetwork waves, gaining
much dynamics from these in the quietest areas where the magnetic
fields have low tension but yet enough to impose slight fibrilarity.
We suspect that slanted internal gravity waves contribute
significantly in this pummeling
   (see Lighthill 1967, pp.~440\,--\,443),
   \nocite{Lighthill1967} 
and together with the primarily vertically growing acoustic shocks
cause intricate and fast-changing interference patterns at granular to
mesogranular scales.  Only the fiercest events punch through to be
individually recognizable directly above their sources.

Our three examples whet the appetite for even more comprehensive
multi-diagnostic solar-atmosphere tomography.  The complexity of the
panel layout in Figures~\ref{fig:grain}\,--\,\ref{fig:exploder} already
demonstrates that the dynamical coupling between photosphere and
chromosphere can only be addressed holistically with diverse
diagnostics.  The complexity of the solar scenes shown in these panels
and their large differences between diagnostics strengthen this
conclusion.  In fact, our data is yet incomplete: we lack the
principal acoustic-event measure of ``acoustic flux'' in this
analysis, requiring Fabry-Perot Doppler mapping of the photosphere at
multiple heights.  Sensitive photospheric magnetic field mapping and
ultraviolet chromospheric image and Doppler diagnostics would also be
welcome -- preferably all at similar or better angular resolution,
cadence, duration, and field size than used here.

In this paper we have limited the discussion to three hand-picked
cases from an early set of tomographic DOT image sequences.  The large
DOT database collected in the meantime (openly available at
\url{http://dotdb.phys.uu.nl/DOT}) now permits wider statistical
study of such various types of events.  Rapid-cadence Fabry-Perot
imaging at the SST promises unprecedented \Halpha\ diagnostics.
Space-mission co-pointing adds EUV diagnostics of the transition
region.  Pertinent numerical simulations including realistic formation
of \Halpha\ are also coming into reach.


\begin{acks}
  The DOT is owned by Utrecht University and located at the Spanish
  Observatorio del Roque de los Muchachos of the Instituto de
  Astrof{\'{\i}}sica de Canarias.  We are deeply indebted to
  V.~Gaizauskas concerning the DOT \Halpha\ filter.  We thank
  M.~Carlsson, \O.~Langangen, S.E.M.~Keek, J.M.D.~Kruijssen and
  A.G. de Wijn for inspiring debates.  This research made much use of
  SST hospitality and of NASA's Astrophysics Data System.  R.J.~Rutten
  thanks the Leids Kerkhoven-Bosscha Fonds and the organizers of the
  SOHO\,19/GONG\,2007 meeting for travel support.
\end{acks}




\end{article}
\end{document}